\documentclass[aps,prd,preprint,superscriptaddress,showpacs,floatfix,nobibnotes,nofootinbib]{revtex4-2}

\usepackage{float}
\usepackage{latexsym}
\usepackage{amsmath}
\usepackage{amssymb}
\usepackage{graphicx}
\usepackage{setspace}
\usepackage{longtable}
\usepackage{enumitem}
\usepackage[bottom]{footmisc}
\usepackage{slashed}
\usepackage{color}

\usepackage{bm}
\usepackage{epsfig}
\newcommand{\bea}{\begin{eqnarray}}
\newcommand{\eea}{\end{eqnarray}}

\newcommand{\nn}{\nonumber}

\begin{document}

\def\be{\begin{equation}}
\def\ee{\end{equation}}
\def\bea {\begin{eqnarray}}
\def\eea {\end{eqnarray}}
\def\bc {\begin{center}}
\def\ec {\end{center}}
\def\nn {\nonumber}
\def\a{\alpha}
\def\b{\beta}
\def\eps {\epsilon}
\def\gm {\gamma}
\def\lm {\Lambda}
\def\mn {\mu\nu}
\def\({\left(}
\def\){\right)}
\def\]{\right]}
\def\sp {\shortparallel}

\newcommand*{\paral}{\stretchrel*{\parallel}{\perp}}
\newcommand{\at}{\tilde{a}}
\newcommand{\bt}{\tilde{b}}
\newcommand{\ct}{\tilde{c}}
\newcommand{\bp}{\bar{p}}

\newcommand{\am}{\textcolor{red}}
\newcommand{\mc}{\textcolor{blue}}
\newcommand{\gk}{\textcolor{ForestGreen}}

\title{The effect of anisotropy on the formation of heavy quarkonium bound states}

\author{Margaret E. Carrington}
\email{carrington@brandonu.ca}
\affiliation{Department of Physics, Brandon University, Brandon, MB R7A 6A9, Canada}
\affiliation{Winnipeg Institute for Theoretical Physics, Winnipeg, MB R3T 2N2, Canada}
\author{Gabor Kunstatter}
\email{gkunstatter@uwinnipeg.ca}
\affiliation{Winnipeg Institute for Theoretical Physics, Winnipeg, MB R3T 2N2, Canada}
\affiliation{Department of Physics, University of Winnipeg, Winnipeg, MB R3M 2E9, Canada}
\author{Arghya Mukherjee}
\email{arbp.phy@gmail.com}
\affiliation{Ramakrishna Mission Residential College (Autonomous), Narendrapur, Kolkata-700103, India}

\date{May 09, 2024}

\begin{abstract}
We  study  the real part of the static potential of a heavy quark-antiquark system in an anisotropic plasma medium. We use a quasi-particle approach where the collective dynamics of the plasma constituents  is described using hard-loop perturbation theory. The parton distribution function is characterized by a set of parameters that can accurately describe the anisotropy of the plasma produced in a heavy ion collision. 
We calculate the potential numerically in strongly anisotropic systems and study the angular dependence of the distortion of the potential relative to the isotropic one.
We  obtain  an analytic expression for the real part of the heavy quark potential in the limit of weak anisotropy using a model that expresses the potential in terms of effective screening masses that depend on the anisotropy parameters and the orientation of the quark-antiquark pair. A 1-dimensional potential is formulated in terms of angle averaged screening masses that incorporate the anisotropy of the medium into a radial coordinate. We solve the corresponding Schr\"odinger equation and show that the magnitude of the binding energy typically increases with anisotropy. Anisotropy can play an important role, especially in states with non-zero angular momentum. This means that the number of bound states that are formed could depend on specific characteristics of the anisotropy of the plasma. Our study suggests that plasma anisotropy plays an important role in the dynamics of heavy quarkonium and motivates further study.

\end{abstract}

\maketitle 

\newpage

\section{Introduction}

The dissociation of heavy quarkonium has been extensively studied as a probe of the properties of quark-gluon plasma. Due to the large mass of the constituent quarks, heavy quarkonia can be treated non-relativistically. One can calculate the heavy quark potential and obtain information about quarkonium binding energies and decay rates by solving a non-relativistic Schr\"odinger equation. In equilibrium plasma there have been many calculations of the heavy quark potential and the properties of heavy quarkonium (see, for example, \cite{Laine:2006ns,Brambilla:2008cx,Carignano:2020gpn,Brambilla:2024tqg}). In recent years progress has been made on the development of methods to go beyond the equilibrium limit \cite{Dumitru:2007hy,Dumitru:2009ni,Nopoush:2017zbu,Dong:2021gnb,Dong:2022mbo,Islam:2022qmj}.

The heavy-quark potential is a complex function and its real/imaginary parts give information on the binding energies / decay widths. In this work we calculate the real part of the potential in a chiral plasma using hard-loop (HL) resummed perturbation theory with an anisotropic distribution function. Our distribution function includes a parameter that produces a spheroidal distortion of the isotropic distribution, as originally introduced in \cite{Romatschke:2003ms}. We include additional anisotropy parameters that give more general momentum distributions and provide a more realistic description of a quark-gluon plasma \cite{Carrington:2021bnk}. 
We find numerical solutions in the chirally symmetric limit. 
When the distribution is spheroidal the potential becomes deeper (more strongly binding) as the anisotropy increases, and the directional dependence of the potential shows that the quark-antiquark pair attracts more strongly when they are aligned in the direction of the anisotropy.
When more realistic anisotropy is introduced this simple picture is destroyed and a much richer structure develops. We  obtain an analytic expression for the potential in the limit of weak anisotropies. 
We reformulate the result in a more useful way using a model that expresses the potential in terms of effective screening masses that depend on the anisotropy parameters and the orientation of the quark-antiquark pair \cite{Dumitru:2009ni}. We  construct a 1-dimensional potential in terms of an angle averaged screening mass that incorporates the anisotropy into a radial coordinate \cite{Dong:2021gnb,Dong:2022mbo,Islam:2022qmj}. Some physics is necessarily lost when this averaging procedure is used, but the huge advantage is that the corresponding Schr\"odinger equation becomes 1-dimensional and is straightforward to solve numerically. We study the dependence of the binding energy on the anisotropy parameters and show that magnitude of the binding energy typically increases with anisotropy. We show that non-spheroidal anisotropy can play an important role, especially in states with non-zero angular momentum. 

This paper is organized as follows. In section \ref{form-sec} we describe the analytic calculation of the potential. 
Sections \ref{pi-tag-sec}-\ref{potential-sec} explain how we calculate the temporal gluon propagator in an anisotropic system and construct the static potential from it. In section \ref{weak-ana} we give an analytic result for the potential in the limit of weak anisotropy. Analytic expressions for the six dressing functions that are used to calculate the potential are given in Appendix \ref{dress-ana}. In section \ref{meff-sec} we discuss the ansatz we use to model the potential in terms of an effective screening mass, and how to reduce it to a function of one radial variable. We also explain the method we use to solve the resulting 1-dimensional Schr\"odinger equation. In section \ref{results-sec} we present some numerical results and in section \ref{conclusions-sec} we make some concluding remarks. 

We use natural units where $\hbar = c =1$. The indices $i,j,k = 1, 2, 3$ and $\mu, \nu = 0, 1, 2, 3$ label, respectively, the Cartesian spatial coordinates and those of Minkowski space. Our metric is mostly minus $g_{\mu\nu} = (1,-1,-1,-1)_{\rm diag}$. 
We use capital letters for four-vectors so that, for example, $P^2=p_0^2-\vec p\cdot\vec p =  p_0^2-p^2$. 

\section{Formalism}
\label{form-sec}
At leading order the heavy quark potential is obtained from the Fourier transform of the temporal component of the retarded gluon effective propagator (${\cal D}_{\mu\nu}$) in the static limit as
\bea
V(\vec r)&=-g^2 C_F \int \frac{d^3p}{(2\pi)^3} \, (e^{i \vec{p}\cdot\vec{r}}-1)\mathcal{D}_{00}(p^0\rightarrow0,\vec p) \,
\label{ms_pot_def}
\eea
where $g$ is the strong coupling constant and $C_F=\frac{4}{3}$ is the quadratic Casimir of the colour  SU(3) group. The inverse of the  effective propagator is related to the gluon polarization tensor ($\Pi_{\mu\nu}$) through the Dyson-Schwinger equation 
\bea
\mathcal{D}_{\mu\nu}^{-1}(p_0,\vec p)&=&(\mathcal{D}_0)_{\mu\nu}^{-1}(p_0,\vec p)+\Pi_{\mu\nu}(p_0,\vec p)\,.
\label{ms_DS}
\eea
We work in covariant gauge where $(\mathcal{D}_0)_{\mu\nu}^{-1}(p_0,\vec p)$ is the inverse of the free gluon propagator 
\bea
\mathcal{D}_0^{\mu\nu}(p_0,\vec p)&=&-\frac{g^{\mu\nu}}{P^2}+(1-\chi)\frac{P^\mu P^\nu}{P^4}\,
\label{ms_free_prop}
\eea
and use Landau gauge which corresponds to the choice $\chi=0$. 
If the medium is isotropic the temporal component of the effective propagator depends only on the magnitude of $\vec p$ and the corresponding potential is spherically symmetric. 
The medium produced in a heavy ion collision is not isotropic, it is strongly anisotropic. The goal of this work is to study how these anisotropies affect the heavy quark potential, and the binding energies and decay rates of quarkonium. The main steps of the work presented in this paper are: 

\begin{enumerate}[label=\Alph*]
\item Calculate the gluon polarization tensor in an anisotropic medium in the HL approximation. 
\item Invert the Dyson equation (\ref{ms_DS}) to find the temporal component of the propagator.

\item Calculate the static potential using equation (\ref{ms_pot_def}). 
\item Represent the anisotropic static potential using an ansatz that models the effect of the anisotropy through effective screening masses. Further simplification is obtained by averaging over angles to get an anisotropic 1-dimensional potential. 
\item Solve the Schr\"odinger equation using the 1-dimensional anisotropic potential.
\end{enumerate}

For the first two parts of the calculation  we consider a medium that can be both  anisotropic and chirally asymmetric.  In the third part we look at plasmas with zero chemical potential and obtain an analytic result for the potential in the limit of weak anisotropy.
For our calculation to be valid perturbation theory must work. The non-relativistic approach that we use to study quarkonium is valid for bound states with large quark masses\footnote{We do not include a non-perturbative string tension term.}.

\subsection{The  polarization tensor}
\label{pi-tag-sec}

In this section we discuss the calculation of the polarization tensor in an anisotropic system using a HL effective theory, following the method of ref. \cite{Carrington:2021bnk}. We start with a brief discussion of the finite temperature calculation for a thermal QED plasma. Time-like axial gauge (TAG) is particularly useful because the gauge condition is imposed in the heat bath rest frame. In TAG only the spatial components of the propagator are non-zero. Our notation for the equilibrium fermion distribution function is 
\bea
n_f(k) =  \frac{1}{e^{\beta(k-\mu)}+1} \text{~~and~~} \bar n_f(k) =  \frac{1}{e^{\beta(k+ \mu)}+1}\, \label{nn-def}
\eea
and the 1-loop photon polarization tensor in the HTL approximation is
\bea
&& \Pi^{ij}(p_0,\vec p) = \Pi^{ij}_{{\rm even}}(p_0,\vec p) + \Pi^{ij}_{{\rm odd}}(p_0,\vec p)\\[2mm]
&& \Pi^{ij}_{{\rm even}}(p_0,\vec p) = 2g^2\int\frac{d^3 k}{(2\pi)^3}\frac{n(k)+\bar n(k)}{k} 
\left(\delta^{ij}+\frac{v^i p^j + p^i v^j}{P\cdot V +i\epsilon}- \frac{P^2 v^i v^j}{(P\cdot V+i\epsilon)^2}\right) \label{pi-even} \\
&& \Pi^{ij}_{{\rm odd}}(p_0,\vec p) =  i g^2 P^2 \epsilon^{ijm} \int\frac{d^3 k}{(2\pi)^3}\frac{n(k) - \bar n(k)}{k^2} 
\frac{(p_0 v^m - p^m)}{(P\cdot V +i\epsilon)^2} \,. \label{pi-odd}
\eea
We use $P\cdot V = p_0-\vec p\cdot\vec v$ and  
$\vec k/\sqrt{k^2+m^2} \approx \hat k \equiv \vec v$ since massless fermions are consistent with the HTL (and HL) approximation.
In a chirally asymmetric system there are two different chemical potentials, for right and left handed fermions. 

To include momentum anisotropy we modify the distribution function in  (\ref{pi-even}, \ref{pi-odd}) using 
\bea
n(k) \to n(\vec k) = C_\xi \, n\big(k H_\xi(\vec v)\big)
\label{norm0}
\eea
and similarly for $\bar n(k)$. The subscript $\xi$ indicates dependence on a set of anisotropy parameters that can be used to construct a distribution that is deformed relative to the isotropic one. The factor $C_\xi$ is a normalization and can be defined in different ways depending on the calculation being done, as explained at the end of this section. 
We can construct a completely general expression for the function $H_\xi(\vec v)$ as a sum of terms that are products of anisotropy parameters and dot products of the vector $\vec v$ with two perpendicular unit vectors. For these unit vectors we use $\hat n_3$ along the beam axis and $\hat n_1$ gives the direction of transverse anisotropy. 
We restrict to functions that satisfy the condition $H_\xi(\vec v) = H_\xi(-\vec v)$ and use an expression of the form\footnote{Note that $\xi_0$ in this paper is defined equal to $\xi_0-1$ in \cite{Carrington:2021bnk}.}
 \bea
 H^2_\xi(\vec v)  &=& (1+\xi_0)
 +\xi _2(\vec n_1\cdot\vec v)^2+\xi _9 (\vec n_3\cdot\vec v)^2
 +\xi _6 (\vec n_1\cdot\vec v)(\vec n_3\cdot\vec v) \label{H-def} \\
     && \hspace*{-1.2cm} 
     +\xi _4 (\vec n_1\cdot\vec v)^4
      + \xi _8 (\vec n_1\cdot\vec v)^3 (\vec n_3\cdot\vec v)
 +\xi _{11}(\vec n_1\cdot\vec v)^2 (\vec n_3\cdot\vec v)^2
     +\xi _{13} (\vec n_1\cdot\vec v)(\vec n_3\cdot\vec v)^3  
     +\xi _{14} (\vec n_3\cdot\vec v)^4 \,. \nonumber
 \eea
The values of the anisotropy parameters $\xi_i$ must be chosen so that $H^2_\xi(\vec v)$ is positive for all orientations of the vector $\vec v$, 
which is equivalent to the requirement that $H_\xi(\vec v)$, and therefore the argument of the distribution function, is real and positive.
For a given choice of the anisotropy parameters, the isotropic distribution is expanded in the direction of $\vec v$ if $H_\xi(\vec v)<1$, and contracted if $H_\xi(\vec v)>1$. 

The polarization tensor has in general 9 independent components. We introduce a complete tensor basis of nine projection operators and decompose the polarization tensor in terms of nine scalar functions. 
From the 3-vectors $\vec p$, $\hat n_1$ and $\hat n_3$ we construct three ortho-normal vectors $(\hat p, ~n_f,~m_F)$:
\bea
&& \hat p = \frac{\vec p}{p} \nn\\
&& n_f = \frac{\tilde n_f}{\sqrt{\tilde n_f\cdot \tilde n_f}} \text{~~with~~} \tilde n_f = n_3 - (n_3\cdot \hat p) \hat p\,~~ \nn\\
&& m_F=\frac{\tilde m_F}{\sqrt{\tilde m_F\cdot \tilde m_F}} \text{~~with~~} \tilde m_F = \tilde m_f - (n_f\cdot\tilde m_f)\,n_f \text{~~and~~} \tilde m_f = n_1 - (n_1 \cdot \hat p) \hat p\,.
\eea
The projection operators are defined as
\bea
&& P_1^{ij} = m_F^i m_F^j \,,~~ P_2^{ij} = \hat p^i \hat p^j \,,~~ P_3^{ij} = n_f^i n_f^j \nn\\
&& P_4^{ij} = \hat p^i n_f^j + n_f^i \hat p^j \,,~~ P_5^{ij} = \hat p^i m_F^j + m_F^i \hat p^j \,,~~ P_6^{ij} =  n_f^i m_F^j + m_F^i n_f^j \nn\\
&& P_7^{ij} =  n_f^i \hat p^j - \hat p^i n_f^j \,,~~P_8^{ij} =  n_f^i m_F^j -  m_F^i n_f^j \,,~~P_9^{ij} =  \hat p^i m_F^j - m_F^i \hat p^j \,
\label{proj-all}
\eea 
and the polarization tensor is decomposed as
\bea
\Pi^{ij} = \sum_{i=1}^6 \pi_i P_i^{ij} +\sum_{i=7}^9 \pi_i P_i^{ij} 
\label{finalPi}
\eea
where we have omitted the functional arguments to shorten the notation. 
The last three projection operators are anti-symmetric in their indices, and the corresponding dressing functions can only be non-zero if the chiral chemical potential $\mu_5\equiv (\mu_R-\mu_L)/2$ is non-zero. 

When $\xi_{i}=0$ we have $H^2_\xi(\vec v)=1$  and the distribution is isotropic and thermal. At zero chemical potential there are only two different non-zero components in equation (\ref{finalPi}) which are the familiar HTL transverse and longitudinal functions
\bea
&& \pi_1=\pi_3= \Pi_T(p_0,\vec p) = m_D^2 \frac{p_0^2}{2p^2} \left(1 - \frac{P^2}{2p_0 p} \ln \left(\frac{p_0+p+i\epsilon}{p_0- p+i\epsilon}\right) \right) \nn \\
&& \pi_2= \Pi_L(p_0,\vec p) = - m_D^2 \frac{p_0^2}{p^2}\left( 1- \frac{p_0}{2p} \ln\left(\frac{p_0+p+i\epsilon}{p_0- p+i\epsilon}\right) \right)
\eea
where the Debye mass is
\bea
m_D^2 = 2 g^2 \int\frac{d^3k}{(2\pi)^3} \frac{1}{k} \big[4 n_f(k)\big] = \frac{g^2 T^2}{3}\,.
\label{debye}
\eea
The corresponding expression in QCD is obtained by replacing $4n_f(k)$ with $2(N_f n_f(k)+N_c n_b(k))$ where $n_b(k) = 1/(e^{k\beta}-1)$, which gives $[m_D^2]_{\rm qcd} = g^2 T^2(N_c+N_f/2)/3$.

Now we discuss the normalization factor $C_\xi$ in equation (\ref{norm0}). We use
a normalization  that leaves the Debye mass invariant under a change of the anisotropy
parameters. We define a parameter that corresponds to the Debye mass (\ref{debye}) in an anisotropic system 
\bea
[m_D^2]_\xi \equiv 8 g^2 \frac{C_\xi}{(2\pi)^3} \int d\Omega \int dk k^2 \frac{1}{k} n_f(kH)\,
\label{int-norm1}
\eea
and $C_\xi$ is determined by requiring $[m_D^2]_\xi  \equiv m_D^2$. 
We make the change of variable $\tilde k = kH$ and find the condition that determines $C_\xi$
\bea
1 = C_\xi \int\frac{d\Omega}{4\pi} \frac{1}{H^2}\,.
\label{norm1}
\eea
The motivation for this choice of normalization is as follows. 
To calculate the binding energies of the quarkonium system we want to set the threshold (the value of the potential as $r$ approaches infinity) to zero. 
The normalization of the distribution should be chosen so that this threshold value is independent of the set of anisotropy parameters that is used. 
This ensures that when we compare the binding energies produced by different distributions we are seeing the effect of the distortion of the distribution that is produced by anisotropy, and not just an overall shift of the potential. 
In sec. \ref{weak-ana} we work in the limit of weak anisotropy and verify analytically that the value of the potential in the limit $r\to \infty$ is independent of the choice of the anisotropy parameters. We have also considered several sets of anisotropy parameters with large values and verified numerically that the threshold of the potential does not change. 

To calculate the real part of the static potential we only need the leading order contributions to the dressing functions in the limit $p_0\to 0$. We will calculate the potential for the special case of a chirally symmetric plasma where only the first six dressing functions are non-zero. 
From the HL integrals (\ref{pi-even}) one can show that the real parts of these dressing functions are even in ${p}_0$ and the imaginary  parts are odd. 
The leading order terms for each dressing function have the form
\bea
    \pi_1&&=m_D^2 \bar \pi_{1R}^{(0)}+\cdots\,,\nn\\
    \pi_2&&=m_D^2 \bar \pi_{2R}^{(2)}\bar{p}_0^2+\cdots\,,\nn\\
     \pi_3&&=m_D^2 \bar \pi_{3R}^{(0)}+\cdots\,,\nn\\
      \pi_4&&=i m_D^2 \bar\pi_{4I}^{(1)}\bar{p}_0+\cdots\,,\nn\\
      \pi_5&&=i m_D^2 \bar\pi_{5I}^{(1)}\bar{p}_0+\cdots\,,\nn\\
       \pi_6&&=m_D^2 \bar \pi_{6R}^{(0)}+\cdots\,
\label{df-exp}
\eea
where $\bar p_0 = p_0/m_D$. 
In the HL effective theory the masses of the plasma partons are neglected and the coefficients on the right side of (\ref{df-exp}) depend only on the direction of the vector $\vec p$. They must be calculated numerically except in special cases. To obtain analytic results we work  in the weak anisotropy limit by expanding in the parameters $\xi_i$. The resulting expressions for the dressing functions are given in Appendix \ref{dress-ana}. 

\subsection{Inversion of the propagator}
\label{4d-inversion-sec}

To calculate the static potential we need to invert the covariant inverse propagator (\ref{ms_DS}) and extract the temporal component. 
To do this we use a covariant basis which can be constructed following the method of ref. \cite{Ghosh:2020sng}. We start with the four 4-vectors $P^\mu$, $b^\mu = (1,0,0,0)$, $c^\mu = (0,\hat n_3)$ and $a^\mu = (0,\hat n_1)$ and construct four perpendicular 4-vectors as follows. First we define $\bt^\mu$ which is perpendicular to $P^\mu$:
\bea
&& V^{\mu\nu} = g^{\mu\nu}-\frac{P^\mu P^\nu}{P^2}
 \nn \\
&& \bt^\mu = V^{\mu\nu}b_\nu=b^\mu-\frac{b\cdot P}{P^2}P^\mu\,.
\eea
Then we construct $\ct$ perpendicular to both $\bt$ and $P^\mu$
\bea
&& B^{\mu\nu}  = \bt^\mu\bt^\nu  \nn \\
&& V_{b}^{\mu\nu} = V^{\mu\nu}-\frac{B^{\mu\nu}}{\bt^2}\,, \nn \\
&& \ct^\mu = V_{b}^{\mu\nu}c_\nu 
 = c^\mu-\frac{c\cdot P}{P^2}P^\mu-\frac{c\cdot \bt}{\bt^2}\bt^\mu\,.
\eea
Finally we define $\at$
\bea
&& C^{\mu\nu}  = \ct^\mu\ct^\nu  \nn \\
&& V_{bc}^{\mu\nu} = V_{b}^{\mu\nu}-\frac{C^{\mu\nu}}{\ct^2}=V^{\mu\nu}-\frac{B^{\mu\nu}}{\bt^2}-\frac{C^{\mu\nu}}{\ct^2}\,\nn \\
&& \at^\mu =V_{bc}^{\mu\nu}a_\nu=a^\mu-\frac{a\cdot P}{P^2}P^\mu-\frac{a\cdot \bt}{\bt^2}\bt^\mu-\frac{a\cdot \ct}{\ct^2}\ct^\mu\,.
\eea
We use the following nine projection operators
\bea
P_1^{\mu\nu}&=& \at^\mu\at^\nu\,\nn \\
P_2^{\mu\nu}&=&\bt^\mu\bt^\nu\,\nn\\
P_3^{\mu\nu}&=&\ct^\mu\ct^\nu\,\nn \\
P_4^{\mu\nu}&=&\bt^\mu\ct^\nu+\bt^\nu\ct^\mu\,\nn \\
P_5^{\mu\nu}&=&\bt^\mu\at^\nu+\bt^\nu\at^\mu\,\nn \\
P_6^{\mu\nu}&=&\ct^\mu\at^\nu+\ct^\nu\at^\mu\,\nn \\
P_7^{\mu\nu}&=&\bt^\mu\ct^\nu-\bt^\nu\ct^\mu\,\nn \\
P_8^{\mu\nu}&=&\bt^\mu\at^\nu-\bt^\nu\at^\mu\,\nn \\
P_9^{\mu\nu}&=&\ct^\mu\at^\nu-\ct^\nu\at^\mu\,
\eea
and decompose the polarization tensor
\bea
\Pi^{\mu\nu} = \sum_{i=1}^9 \Pi_i P_i^{\mu\nu} \,
\label{PI-def}
\eea
where we have supressed the functional arguments. 
Now we write the covariant gauge propagator in the form
\bea
{\cal D}^{\mu\nu}(p_0,\vec p) = -\frac{g^{\mu\nu}}{P^2}+(1-\chi)\frac{P^\mu P^\nu}{P^4} - \sum {\cal C}_i(p_0,\vec p) P_i^{\mu\nu}
\label{D-cov}
\eea
where the ${\cal C}_i$ are a set of scalar coefficients that can be expressed in terms of the components of the self-energy using equations (\ref{ms_DS}, \ref{ms_free_prop}, \ref{PI-def}, \ref{D-cov}) by solving 
${\cal D}^{-1}_{\mu\lambda}{\cal D}^{\lambda\nu} = g_\mu^{~\nu}$. 
We also need to rewrite the components of the polarization tensor in the 4-dimensional covariant basis, denoted $\Pi_i$, in terms of the components in the time-like-axial gauge, denoted $\pi_i$, which have been calculated in the HL approximation in sec. \ref{pi-tag-sec} and Appendix \ref{dress-ana}. Comparing equations (\ref{finalPi}, \ref{PI-def}) it is straightforward to show
\begin{align}
 \Pi_1&=-\frac{p_{\perp}^2}{p_y^2}\pi_1\,,\,\Pi_2=-\frac{P^4}{p_0^2 p^2}\pi_2\,,\, \Pi_3=-\frac{p^2}{p_\perp^2}\pi_3\,,\,\Pi_4=\frac{P^2}{p_0\sqrt{p_\perp^2}}\pi_4\,,\,\Pi_5=\frac{P^2\sqrt{p_\perp^2}}{p_0 p \sqrt{p_y^2}}\pi_5\,,\,\nn\\
 \Pi_6&=-\frac{p}{\sqrt{p_y^2}}\pi_6\,,\,\Pi_7=-\frac{P^2}{p_0\sqrt{p_\perp^2}}i\pi_7\,,\,\Pi_8=\frac{P^2\sqrt{p_\perp^2}}{p_0 p \sqrt{p_y^2}}i\pi_9\,,\,
 \Pi_9=-\frac{p}{\sqrt{p_y^2}}i\pi_8
      \end{align}
where we have defined $p_\perp^2=p_x^2+p_y^2$.
The final expression  for the temporal component of the propagator is
 \bea
\mathcal{D}^{00}(p_0,\vec p)
&=&-\Big[(P^2-\pi_3 )(P^2-\pi_1 )- (\pi^2_6+\pi^2_8)\Big]/\Big[(P^2-\pi_1)(P^2-\frac{P^2}{p_0^2}\pi_2)(P^2-\pi_3)\nn\\&&~~~-\frac{P^2}{p_0^2}(\pi_4^2+\pi_7^2)(P^2-\pi_1)-(\pi_6^2+\pi_8^2)(P^2-\frac{P^2}{p_0^2}\pi_2)-\frac{P^2}{p_0^2}(\pi_5^2+\pi_9^2)(P^2-\pi_3)\nn\\&&~~~-2\frac{P^2}{p_0^2}(\pi_4\pi_5\pi_6-\pi_6\pi_7\pi_9+\pi_5\pi_7\pi_8+\pi_4\pi_8\pi_9)\Big]\,.
\label{prop-static}
\eea 

Next we discuss the static limit of the temporal propagator in equation (\ref{prop-static}). 
For a chirally symmetric plasma $\pi_7=\pi_8=\pi_9=0$ and equations (\ref{df-exp}, \ref{prop-static}) give an expression for the static temporal propagator. 
We use the notation $\bar p = p/ m_D$ and $\hat p = \vec p /p$ and define $\mathcal{\bar{D}}_{00}(\bar{p},\hat p) = m_D^2 \lim_{p_0\to 0} \mathcal{{D}}_{00}(p_0,\vec p)$. We will also use the notation $\bar\pi_i = \pi_i/m_D^2$ for each component of the polarization tensor. 
The temporal component of the propagator in the static limit is  
\begin{align}
  \mathcal{\bar{D}}_{00}(\bar{p},\hat p)&=\frac{(\bar{p}^2+\bar{\pi}_{1R}^{(0)})(\bar{p}^2+\bar{\pi}_{3R}^{(0)})-(\bar{\pi}_{6R}^{(0)})^2}{a(\bar{p}^2)^3+b(\bar{p}^2)^2+c\bar{p}^2+d}
\label{doo-x}
\end{align}
where
\begin{align}
    a&=1\,,\nonumber\\
    b&=\bar{\pi}_{1R}^{(0)}-\bar{\pi}_{2R}^{(2)}+\bar{\pi}_{3R}^{(0)}\,,\nonumber\\
    c&=-\bar{\pi}_{1R}^{(0)} \bar{\pi}_{2R}^{(2)}-\bar{\pi}_{2R}^{(2)}\bar{\pi}_{3R}^{(0)}+\bar{\pi}_{3R}^{(0)}\bar{\pi}_{1R}^{(0)}-(\bar{\pi}_{4I}^{(1)})^2-(\bar{\pi}_{5I}^{(1)})^2-(\bar{\pi}_{6R}^{(0)})^2\,,\nonumber\\
    d&=-\bar{\pi}_{1R}^{(0)}\bar{\pi}_{2R}^{(2)}\bar{\pi}_{3R}^{(0)}+2\bar{\pi}_{4I}^{(1)}\bar{\pi}_{5I}^{(1)}\bar{\pi}_{6R}^{(0)}-\bar{\pi}_{1R}^{(0)}(\bar{\pi}_{4I}^{(1)})^2-\bar{\pi}_{3R}^{(0)}(\bar{\pi}_{5I}^{(1)})^2+\bar{\pi}_{2R}^{(2)}(\bar{\pi}_{6R}^{(0)})^2\,.
\label{for-doo-x}
   \end{align}
In a spheroidally symmetric system, which can be described with one anisotropy parameter, $\pi_{i}=0$ for $i\ge 5$ and the static temporal propagator takes the simple form
\bea
\mathcal{\bar{D}}^{\rm spheroidal}_{00}(\bar p,\hat p)  = \frac{ \big(\bar p^2 + \bar\pi^{(0)}_{3R} \big)}
{\big(\bar p^2- \bar\pi^{(2)}_{2R}\big) \big(\bar p^2+ \bar \pi^{(0)}_{3R}\big)+\big(\bar\pi^{(1)}_{4I}\big)^2}\,.
\label{prop-mike}
\eea
In the isotropic limit $\bar\pi^{(1)}_{4I}=0$ and $\bar \pi_{2R}^{(2)} = -1$ so  we recover the Debye screening form 
\bea
\mathcal{\bar{D}}^{\rm iso}_{00}(\bar p) = \frac{1}{\bar p^2+1}\,.
\label{prop-iso}
\eea

The denominator of (\ref{doo-x}) is a cubic equation in $\bar p^2$ and therefore has three roots. 
As explained under equation (\ref{df-exp}), the coefficients of the polarization tensor depend only on the direction of the momentum vector $\vec p$ and therefore we denote the roots of the denominator $\bar m_i^2(\hat p)$ with $i\in(1,3)$. One can  write the static propagator in terms of these roots as
\bea
\mathcal{\bar{D}}_{00}(\bar{p},\hat p) = 1+ \sum_{i=1}^3 \frac{{\cal A}_i(\hat p)}{\bar p^2+\bar m_i^2(\hat p)} 
\eea
where
\bea
{\cal A}_1(\hat p) = \frac{\bar m_1^2}{(\bar m_1^2 - \bar m_2^2)(\bar m_1^2 - \bar m_3^2)}
\left[
\bar m_1^2\big(\bar \pi_{1R}^{(0)}+\bar \pi_{3R}^{(0)}\big) - \bar m_1^4
+ \big(\bar \pi_{6R}^{(0)}\big)^2
- \bar \pi_{1R}^{(0)}\bar \pi_{3R}^{(0)}
\right]
\eea
and ${\cal A}_2(\hat p)$ and ${\cal A}_3(\hat p)$ are obtained from ${\cal A}_1(\hat p)$ by cyclically rotating $(\bar m_1,\bar m_2,\bar m_3)$. The roots $(\bar m_1,\bar m_2,\bar m_3)$ are dimensionless mass-like parameters that depend only on $\hat p$ but they have a complicated general form and can be complex.

\subsection{The static potential}
\label{potential-sec}

The static potential is obtained from the temporal component of the gluon propagator using equation (\ref{ms_pot_def}). 
We use 
\bea
&& \hat{p} =(\sin\theta_p\cos\phi_p,\sin\theta_p\sin\phi_p,\cos\theta_p)\,\nn\\
&& \vec r = r (\sin\theta\cos\phi,\sin\theta\sin\phi,\cos\theta)\,\label{r-angles}
\eea
and we define $\alpha=(g^2 C_F)/(4\pi)$ and $x_p=\cos\theta_p$ and use the dimensionless radial variable $\bar{r}= r m_D$. Equation (\ref{ms_pot_def}) then takes the form
\bea
V(\bar r,\theta,\phi)
&=& -\frac{4\pi \alpha m_D}{(2\pi)^3}\int_{-1}^1 dx_p \, \int_0^\infty d\bar{p}\,\bar{p}^2 \int_0^{2\pi} \,d\phi_p \, (e^{i\, \vec{p}\cdot\vec{r}}-1) \mathcal{\bar{D}}_{00}(\bar{p},x_p,\phi_p) \,.
\label{pot-anio}
\eea
In the isotropic limit the propagator (\ref{prop-iso}) gives the Debye screened Yukawa potential
\bea
V_{\rm iso}(\bar r) = -\alpha m_D \left(1+\frac{e^{-\bar r}}{\bar r}\right)\,.
\label{V-iso}
\eea
In the limit $\bar r \to 0$ the isotropic potential reduces to the Coulomb potential which verifies that medium effects disappear at very small distances. 
In the asymptotic limit $\lim_{\bar r \to \infty}V_{\rm iso}(\bar r) = -\alpha m_D$.

The potential in an anisotropic medium can be obtained from equations (\ref{doo-x}, \ref{for-doo-x}, \ref{pot-anio}). For arbitrary sets of anisotropy parameters the dressing functions must be calculated numerically and therefore we are also only able to find the potential numerically. In section \ref{anio-strong} we show some results for the potential in strongly anisotropic systems. 

\subsection{Weak anisotropy}
\label{weak-ana}
It is useful to study the limit of weak anisotropy where we have analytic results for the dressing functions (see Appendix \ref{dress-ana}). It will be convenient to work in the coordinate system defined by the vectors
\begin{align}
    \hat{r}&=(0,0,1)\,,\nn\\
    \hat{n}_1&=(\sin\phi,\cos\theta \cos\phi,\sin\theta\cos\phi)\,,\nn\\
    \hat{n}_3&=(0,-\sin\theta, \cos\theta)\,.
    \label{ref_frame_pot}
\end{align}
In the weakly anisotropic limit the static propagator has the form
\begin{align}
 \mathcal{\bar{D}}_{00}(\bar p,x,\phi)&=\mathcal{\bar{D}}_{00}^{(1)}(\bar p,x,\phi)-\frac{1}{\left(\bar{p}^2+1\right)^2}(C_\xi-1)\,
\label{frame_ind_propagator_C}
\end{align}
where the factor $C_\xi$ is defined in equation (\ref{norm1}). 
To linear order in the anisotropy parameters the normalization $C_\xi$ is 
\bea
C_\xi=1+\sum_i\xi_i g_i\,
\label{CxiDxi}
\eea
with
 \bea
(g_0,g_2,g_4,g_9,g_{11},g_{14})=\left(1,\frac{1}{3},\frac{1}{5},\frac{1}{3},\frac{1}{15},\frac{1}{5}\right)\,.
\label{littleg}
\eea
Equations (\ref{doo-x}, \ref{for-doo-x}, \ref{dress-ana-real}, \ref{dress-ana-imag}) give 
 \begin{align} \mathcal{\bar{D}}_{00}^{(1)}(\bar p,x,\phi)&= \frac{1}{\bar{p}^2+1}+\frac{\xi_0}{\left(\bar{p}^2+1\right)^2}
+\frac{\xi _9 \left(2-3 \left(\hat{p}\cdot \hat{n}_3\right){}^2\right)}{3 \left(\bar{p}^2+1\right)^2}+\frac{\xi _2 \left(2-3 \left(\hat{p}\cdot \hat{n}_1\right){}^2\right)}{3 \left(\bar{p}^2+1\right)^2}\nn\\
    &+\frac{\xi _4 \left(20 \left(\hat{p}\cdot \hat{n}_1\right){}^4-30 \left(\hat{p}\cdot \hat{n}_1\right){}^2+9\right)}{15 \left(\bar{p}^2+1\right)^2}+\frac{\xi _8 \left(\hat{p}\cdot \hat{n}_1\right) \left(4 \left(\hat{p}\cdot \hat{n}_1\right){}^2-3\right) \left(\hat{p}\cdot \hat{n}_3\right)}{3 \left(\bar{p}^2+1\right)^2}\nn\\
    &+\frac{\xi _{11} \left(5 \left(4 \left(\hat{p}\cdot \hat{n}_3\right){}^2-1\right) \left(\hat{p}\cdot \hat{n}_1\right){}^2-5 \left(\hat{p}\cdot \hat{n}_3\right){}^2+3\right)}{15 \left(\bar{p}^2+1\right)^2}-\frac{\xi _6 \left(\hat{p}\cdot \hat{n}_1\right) \left(\hat{p}\cdot \hat{n}_3\right)}{\left(\bar{p}^2+1\right)^2}\nn\\
    &+\frac{\xi _{13} \left(\hat{p}\cdot \hat{n}_1\right) \left(\hat{p}\cdot \hat{n}_3\right) \left(4 \left(\hat{p}\cdot \hat{n}_3\right){}^2-3\right)}{3 \left(\bar{p}^2+1\right)^2}+\frac{\xi _{14} \left(20 \left(\hat{p}\cdot \hat{n}_3\right){}^4-30 \left(\hat{p}\cdot \hat{n}_3\right){}^2+9\right)}{15 \left(\bar{p}^2+1\right)^2}\,.
    \label{frame_ind_propagator}
 \end{align}
Equations (\ref{frame_ind_propagator_C}, \ref{frame_ind_propagator}) are valid only in the limit of weak anisotropy, to linear order in the anisotropy parameters, but to simplify the notation we do not introduce an additional subscript to indicate this. 

To find the potential in the limit of weak aniostropy we substitute (\ref{frame_ind_propagator_C}, \ref{frame_ind_propagator}) into (\ref{pot-anio}) and perform the angular integrals. 
The result for the potential has the form
\bea
&& V_{\rm aniso}(\bar r,\theta,\phi) 
= V_{\rm aniso}^{(1)}(\bar r,\theta,\phi)
 - \frac{\alpha m_D}{2}(1-e^{-\bar{r}})(C_\xi-1)\,
\label{V-anio-N}
\eea
where
\bea
&& V_{\rm aniso}^{(1)}(\bar r,\theta,\phi) = -\frac{\alpha m_D}{\bar r}e^{-\bar r}(1-\sum_i \xi_i f_i(\bar r,\theta,\phi))-\alpha \bar m_D \label{V-anio}
\eea
with
\bea
&& \bar m_D = m_D(1-\sum_i \xi_i h_i) \,.
\label{h-def}
\eea
The non-zero $h_i$ coefficients are 
\bea
(h_0,h_2,h_4,h_9,h_{11},h_{14})=\left(\frac{1}{2},\frac{1}{6},\frac{1}{10},\frac{1}{6},\frac{1}{30},\frac{1}{10}\right)\,.
\label{littleh}
\eea
The $f_i$ coefficients depend on the orientation of the position vector ($\vec r$) but can be written in a frame independent way.
The coefficients $f_0$, $f_9$, $f_2$, $f_6$ have the fairly simple form 
\bea
&& f_0= -\frac{\bar{r}}{2}\nn\\
&& f_9 =\frac{1}{6\bar{r}^2}\Big[3\left(\bar{r}^3+3 \bar{r}^2+6 \bar{r}-6 e^{\bar{r}}+6\right)\left(\hat{r}\cdot \hat{n}_3\right){}^2-2\bar{r}^3-3 \bar{r}^2-6 \bar{r}+6 \left(e^{\bar{r}}-1\right)\Big]~, \nn \\
&& f_2 = \frac{1}{6\bar{r}^2}\Big[3\left(\bar{r}^3+3 \bar{r}^2+6 \bar{r}-6 e^{\bar{r}}+6\right)\left(\hat{r}\cdot \hat{n}_1\right){}^2-2\bar{r}^3-3 \bar{r}^2-6 \bar{r}+6 \left(e^{\bar{r}}-1\right)\Big]~,\nn \\
&& f_6 = \frac{1}{2 \bar{r}^2} \left(\bar{r}^3+3 \bar{r}^2+6 \bar{r}-6 e^{\bar{r}}+6\right)\left(\hat{r}\cdot \hat{n}_1\right)\left( \hat{r}\cdot \hat{n}_3\right)~.
\label{easy}
\eea
As expected from the structure of equation (\ref{H-def}), the coefficient $f_2$ can be obtained from $f_9$ by replacing $\hat{r}\cdot \hat{n}_3$ with $\hat{r}\cdot \hat{n}_1$. 
The frame independent expressions for the  coefficients $f_4$, $f_8$ and $f_{11}$  are 
\bea
f_4&=&\frac{1}{\bar{r}^4}\Big[-\frac{2}{3}\left(\hat{r}\cdot \hat{n}_1\right){}^4 f_4^{(4,0)}+\left(\hat{r}\cdot \hat{n}_1\right){}^2f_4^{(2,0)}+\frac{1}{10}f_4^{(0,0)}\Big]\nn\\
     f_8&=&\frac{1}{\bar{r}^4}\Big[\frac{1}{2}\left(\hat{r}\cdot \hat{n}_1\right)\left(\hat{r}\cdot \hat{n}_3\right)f_8^{(1,1)}-\frac{2}{3}\left(\hat{r}\cdot \hat{n}_1\right)^3\left(\hat{r}\cdot \hat{n}_3\right)f_8^{(3,1)}\Big]\nn\\
     f_{11}&=&\frac{1}{\bar{r}^4}\Big[\frac{1}{6}\left(\hat{r}\cdot \hat{n}_1\right){}^2 f_{11}^{(2,0)}-\frac{2}{3}\left(\hat{r}\cdot \hat{n}_1\right){}^2 \left(\hat{r}\cdot \hat{n}_3\right){}^2 f_{11}^{(2,2)}+\frac{1}{6}\left(\hat{r}\cdot \hat{n}_3\right){}^2 f_{11}^{(0,2)}+\frac{1}{30}f_{11}^{(0,0)}\Big]\nn \\
       \label{hard}
\eea
where
\bea
f_4^{(4,0)}&=&f_8^{(3,1)}=f_{11}^{(2,2)}=\bar{r}^5+10 \bar{r}^4+55 \bar{r}^3+15 \left(e^{\bar{r}}+13\right) \bar{r}^2+420 \bar{r}-420 \left(e^{\bar{r}}-1\right)\nn\\
f_4^{(2,0)}&=&f_8^{(1,1)}=f_{11}^{(2,0)}=f_{11}^{(0,2)}=\bar{r}^5+7 \bar{r}^4+34 \bar{r}^3+6 \left(e^{\bar{r}}+19\right) \bar{r}^2+240 \bar{r}-240 \left(e^{\bar{r}}-1\right)\nn\\
f_4^{(0,0)}&=&=f_{11}^{(0,0)}=-3\bar{r}^5-10 \bar{r}^4-40 \bar{r}^3-120 \bar{r}^2-240 \bar{r}+240 \left(e^{\bar{r}}-1\right)\,.
\label{hard2}
\eea
The coefficients $f_{14}$ and $f_{13}$ can be obtained from $f_4$ and $f_8$ respectively by interchanging $\hat{r}\cdot \hat{n}_3$ and $\hat{r}\cdot \hat{n}_1$. 

We consider several limits of the anisotropic potential in equation (\ref{V-anio-N}). 
It is straightforward to see that in the limit $\xi_i\to 0$ the anisotropic potential  reduces to the isotropic result in equation (\ref{V-iso}). 
From equations (\ref{easy}, \ref{hard}, \ref{hard2}) one can show that the factor $e^{-\bar r} f_i/\bar r $ goes to zero in the limit $\bar r\to\infty$ for all values of $i$. 
The threshold of the potential (\ref{V-anio-N}) is therefore $-\alpha[\bar m_D + m_D (C_\xi-1)/2]$ and from (\ref{littleg}, \ref{h-def}, \ref{littleh}) it is easy to verify this is the isotropic threshold value $-\alpha m_D$. 
One can also show that the anisotropic potential (\ref{V-anio-N}) reduces to the Coulomb potential in the 
limit $\bar r\to 0$. To see this we use equations (\ref{littleh}, \ref{easy}, \ref{hard}, \ref{hard2}) to show that for $i\in\{0,2,4,9,11,14\}$ the functions $f_i(\bar r,\theta,\phi)$ and constants $h_i$ satisfy the relation 
\bea
\lim_{\bar r \to 0}f_i(\bar r,\theta,\phi)  = -\bar r h_i\,
\label{h-f-rel}
\eea
and for $i\in\{6,8,13\}$
\bea
\lim_{\bar r \to 0}f_i(\bar r,\theta,\phi)  = {\cal O}(\bar r^2)\,.
\label{h-f-rel-2}
\eea

\subsection{Ansatz for the anisotropic potential}
\label{meff-sec}

We would like to solve the 3-dimensional Schr\"odinger equation constructed from the anisotropic potential in equation (\ref{V-anio-N}). Although this calculation is straightforward to formulate numerically it is difficult to carry out in practice.  We therefore concentrate on a simpler approach  \cite{Dumitru:2009ni,Dong:2021gnb,Dong:2022mbo,Islam:2022qmj} which will let us assess the importance of the anisotropy parameters we have introduced. The basic idea is to introduce an ansatz for the anisotropic potential that allows us to model the effects of the anisotropy in terms of an effective screening mass that is position dependent, and then work with an angle averaged screening mass that depends on only the radial coordinate.  

\subsubsection{Screening masses}

To begin we consider the simple ansatz 
\bea
\tilde V_{\rm ansatz}(\bar r,\theta,\phi) = -\alpha \bar m_D - \frac{\alpha}{r} e^{-r\tilde m_D(\bar r,\theta,\phi)}\,
\label{ansatz2}
\eea
where $\bar m_D$ is the constant mass scale given in equations (\ref{h-def}, \ref{easy}, \ref{hard}) and $\tilde m_D$ is an effective screening mass that depends on  the anisotropy parameters and position. The physical motivation for the ansatz in equation (\ref{ansatz2}) is easy to understand. The effective screening mass should be the inverse of the length scale where screening effects become important in an anisotropic medium. It should therefore be the inverse of a screening length scale, denoted $r_s$, defined through an equation of the form
\bea
\frac{|r \tilde V_{\rm ansatz}(r)|_{r\to 0}}{|r \tilde V_{\rm ansatz}(r)|_{r\to r_s}} = \frac{1}{e}\,.
\label{screen-length}
\eea
The numerator in equation (\ref{screen-length}) is $-\alpha$. 
If we drop the constant in (\ref{ansatz2}), which is equivalent to setting the threshold to zero,  the denominator is $-\alpha e^{-\tilde m r_s}$ and eq. (\ref{screen-length}) gives $r_s\sim 1/\tilde m_D$. 

At very small distances we should find that medium effects disappear and we recover the Coulomb limit. 
This physical property of the static potential is not satisfied by the ansatz (\ref{ansatz2}) which gives
\bea
\lim_{r\to 0} \tilde V_{\rm ansatz}(\vec r) = -\frac{\alpha}{r}-\alpha(\bar m_D-\tilde m_D)\,.
\eea
We use a modified ansatz that preserves the Coulomb limit \cite{Dong:2021gnb}
\bea
V_{\rm ansatz}(\vec r) = -\alpha \bar m_D - \frac{\bar m_D}{\tilde m_D}\frac{\alpha}{r} e^{-r \tilde m_D}+\frac{\alpha}{r}\left(\frac{\bar m_D}{\tilde m_D}-1\right)\,
\label{ansatz1}
\eea
where
\bea
&& \tilde m_D \equiv m_D\big(1+ \sum_i \xi_i c_i(\bar r, \theta, \phi)\big)\,.
\label{mass2}
\eea 
The potential (\ref{ansatz1}) reduces to the isotropic form when the anisotropy parameters are set to zero and gives the Coulomb potential in the limit that $r$ goes to zero. 
The coefficients $c_i$ are determined by matching (\ref{ansatz1}) to our result (\ref{V-anio-N}) with the threshold values subtracted.
This gives
\bea
c_i(\bar r,\theta,\phi) 
= \frac{2f_i(\bar r,\theta,\phi) + 2h_i(e^{\bar r} -1) + g_i{\bar r} }{2(1-e^{\bar r}+\bar r)}\,
\label{c-coefs}
\eea
where the coefficients $f_i(\bar r,\theta,\phi)$ and $h_i$ and $g_i$ are given in equations (\ref{littleg}, \ref{littleh}, \ref{easy}, \ref{hard}, \ref{hard2}).

\subsubsection{Angle averaged effective masses}

To avoid the numerical difficulties of solving a 3-dimensional differential equation we use an angle averaged screening mass $\mathcal{M}_{lm}$ defined as 
\begin{align}
   \mathcal{M}_{lm}(\xi_i)&=\int d\Omega Y_{lm}^*(\theta,\phi) \tilde{m}_D(\xi_i,\theta,\phi) Y_{lm}(\theta,\phi) 
\label{curleyM}
\end{align}
which depends only on the anisotropy parameters $\xi_i$. 
These screening masses are 
\begin{align}
\mathcal{M}_{lm}&= M_{lm} +
m_D(C_\xi-1)\left(\frac{1}{3}-\frac{1}{\bar r} -\frac{\bar r}{36}\right) \nn \\
  M_{00}&=m_D\Bigg[1-\frac{\xi_0}{2}-\frac{\xi_{9}}{6}-\frac{\xi_{2}}{6}-\frac{\xi_{11}}{30}-\frac{\xi_{4}}{10}-\frac{\xi_{14}}{10}\Bigg]\nn\\
  M_{10}&=m_D\Bigg[1-\frac{\xi_0}{2}-\frac{\xi_{9}}{10}-\frac{\xi_{2}}{5}-\frac{\xi_{11}}{35}-\frac{9 \xi_{4}}{70}-\frac{3 \xi_{14}}{70}+\bar{r} \left(-\frac{\xi_{11}}{1575}-\frac{4 \xi_{14}}{525}+\frac{\xi_{2}}{225}+\frac{2 \xi_{4}}{525}-\frac{2 \xi_{9}}{225}\right)\Bigg]\nn \\
   M_{11}&=m_D\Bigg[1-\frac{\xi_0}{2}-\frac{\xi _9}{5}-\frac{3 \xi _2}{20}-\frac{\xi _{11}}{28}-\frac{3 \xi _4}{35}-\frac{9 \xi _{14}}{70}+\bar{r}\left(-\frac{\xi _2}{450}+\frac{\xi _9}{225}+\frac{\xi _{11}}{3150}+\frac{2 \xi _{14}}{525}-\frac{\xi _4}{525}\right) \Bigg]\nn \\
   M_{1-1}&=M_{11}\,.\label{M-ave}
\end{align}
These expressions show that the  anisotropy parameters that break the spheroidal symmetry of the distribution can play an important role, especially for states with non-zero angular momentum quantum numbers.

\subsubsection{Schr\"odinger equation and its solutions}

It is reasonably straightforward to solve the Schr\"odinger equation using the angle averaged effective mass $\mathcal{M}_{lm}$ in the ansatz (\ref{ansatz1}) to find the wavefunction and binding energy for the state with quantum numbers $(l,m)$.
The numerical difficulty is vastly reduced because the differential equation that must be solved is 1-dimensional, and the physical effects of the anisotropy are packaged into the screening masses.
The validity of the procedure is analysed in \cite{Dong:2021gnb} and verified in \cite{Islam:2022qmj}. 

The Schr\"odinger equation for a potential that depends on one radial coordinate can be separated in the usual way by writing 
$\psi(\vec r) = Y_{l0}(\theta,\phi) R_{nl}(r)$ and $u_{nl}(r) = r R_{nl}(r)$ so that the differential equation to be solved has the form
\bea
&& -\frac{\hbar^2}{2m} \left(\frac{d^2 u(r)}{dr^2} - \frac{l(l+1)u(r)}{r^2} \right) + V_{lm}(r) u(r) = E u(r)\,.
\label{schoe}
\eea
We use $2m=M_Q$ where $M_Q$ is the mass of the heavy quark and scale variables using the Bohr radius $a=4\pi/(g^2 C_F M_Q)$ defining $\tilde r = r/a$ and $\tilde V = M_Q a^2 V$ and $\tilde E = M_Q a^2 E$. 
In terms of these variables the Schr\"odinger equation has the form
\bea
-\frac{d^2 u(\tilde r)}{d\tilde r^2} + \tilde V_{\rm eff}(\tilde r) u (\tilde r)  = \tilde E u (\tilde r) \label{my-sch}
\eea
with
\bea
\tilde V_{\rm eff}(\tilde r)  = \left(\tilde V(\tilde r)  + \frac{l(l+1)}{\tilde r^2}\right)\,.
\label{my-sch2}
\eea

We use the method of \cite{Falkensteiner, Lucha:1998xc} to solve equations (\ref{my-sch}, \ref{my-sch2}). The calculation uses the fact that $\tilde V(\tilde r)$ must be less singular than $-1/\tilde r^2$ for the energy eigenvalue to be bounded from below, and that if $\tilde r^2 V(\tilde r)$ is analytic the solution of the Schr\"odinger equation must have the form $u(\tilde r) \propto \tilde r^{l+1}[1+{\cal O}(\tilde r)]$ at small $\tilde r$. We can therefore solve the second order  differential equation (\ref{my-sch}) with the boundary conditions $u(\tilde r_{\rm min}) = \tilde r_{\rm min}^{l+1}$ and $u'(\tilde r_{\rm min}) = (l+1)\tilde r_{\rm min}^{l}$ where $\tilde r_{\rm min}$ is a small number that must be greater than zero for numerical reasons (we use $10^{-3}$). 
The idea of the method is to find the true eigenvalue by performing a systematic search for the eigenvalue that produces a normalizable eigenfunction. The state is identified from the number of nodes. 

\section{Results}
\label{results-sec}

We define $\hat V = V/(g^2 C_F m_D)$ so that $\hat V_{\rm iso} = -e^{-\bar r}/(4\pi \bar r)$ and $\hat V_{\rm vacuum} = -1/(4\pi \bar r)$.
In all of the cases we discuss, when the value of an anisotropy parameter is not given it is set to zero. The angles $\theta$ and $\phi$ are defined in equation (\ref{r-angles}) and we use $x=\cos(\theta) = \cos(\hat n_3 \cdot \hat r)$. 

\subsection{Strong anisotropy}
\label{anio-strong}

\begin{figure}[H]
\begin{center}
\includegraphics[width=8cm]{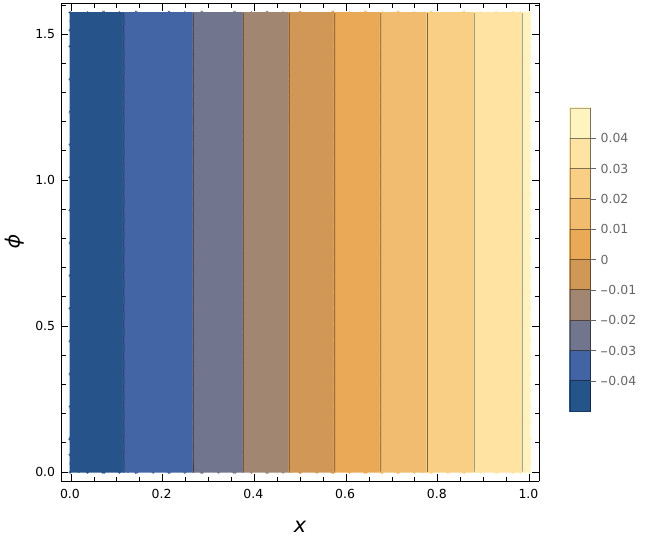}
\includegraphics[width=8cm]{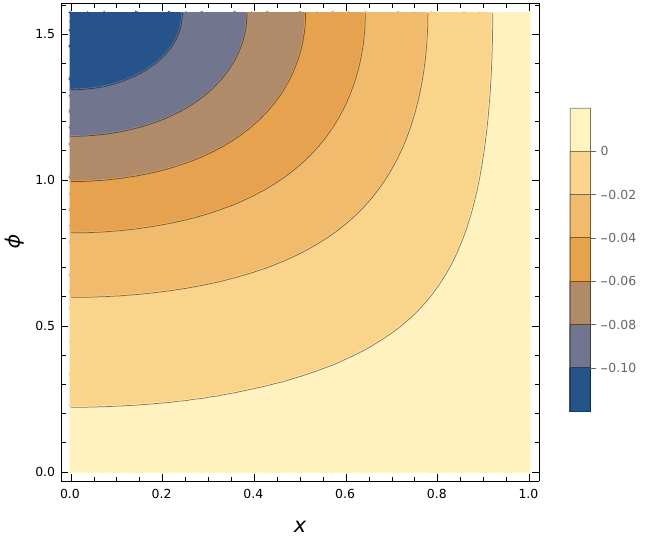}
\end{center}
\caption{Contour plots of the correction to the isotropic potential divided by the vacuum potential at $\bar r=0.8$. The left panel shows $\xi_9=10$ and the right panel is $\xi_9=10$ and $\xi_2=8$.  \label{plot2}}
\end{figure}

In fig. \ref{plot2} we show the correction to the isotropic potential divided by the vacuum potential as a function of $x=\cos(\theta)$ and $\phi$ at $\bar r=0.8$ for two different sets of parameters. The left panel shows that when the distribution function has spheroidal symmetry the potential depends only on the polar angle, and is deepest when the quark-antiquark pair are aligned with the direction of the anisotropy. The right panel shows the dependence of the potential on the azimuthal angle when one additional anisotropy  parameter is introduced. 
\par

\begin{figure}[H]
\begin{center}
\includegraphics[width=12cm]{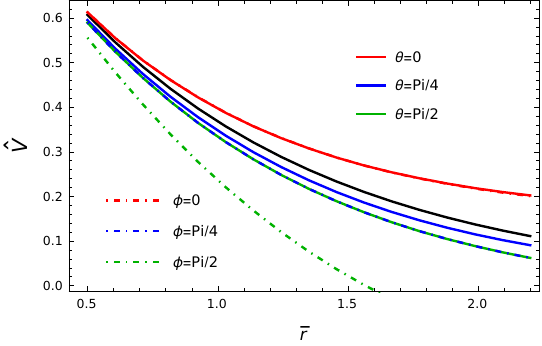}
\end{center}
\caption{The potential divided by the vacuum potential for three values of $\theta$ at $\phi=\pi/4$ and three values of $\phi$ at $\theta=\pi/2$ using the anisotropy parameters $(\xi_0,\xi_2,\xi_4,\xi_6,\xi_8,\xi_9,\xi_{11},\xi_{13},\xi_{14}) = (5,8,37,-21,21,8,15,-28,40)$.  \label{plot1}}
\end{figure}

In fig. \ref{plot1} we show the potential divided by the vacuum potential for three values of $\theta$ at $\phi=\pi/4$ and three values of $\phi$ at $\theta=\pi/2$. The anisotropy parameters used are $(\xi_0,\xi_2,\xi_4,\xi_6,\xi_8,\xi_9,\xi_{11},\xi_{13},\xi_{14}) = (5,8,37,-21,21,8,15,-28,40)$. 
\par

\subsection{Weak anisotropy}

\begin{figure}[H]
\begin{center}
\includegraphics[width=12cm]{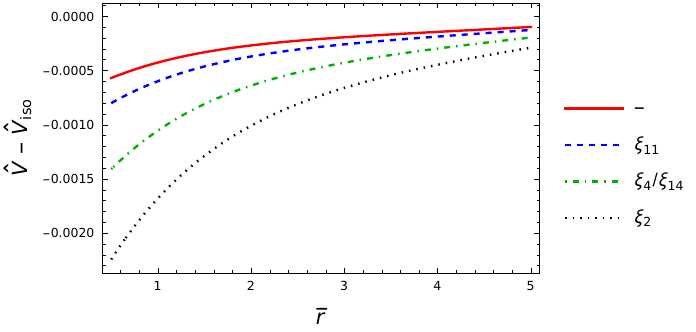}
\end{center}
\caption{The angle averaged ansatz for the anisotropic potential minus the isotropic potential using ${\cal M}_{00}$. In all cases $\xi_9=0.95$, the labelled parameter(s) is $0.8$, and all other parameters are set to zero. \label{plot3}}
\end{figure}

We consider five different sets of parameters. 
In the first set we use $\xi_9=0.95$ and all other parameters set to zero. 
The next four sets have $\xi_9=0.95$, one of the parameters ($\xi_2$, $\xi_4$, $\xi_{14}$, $\xi_{11}$) set to $0.8$, and all others set to zero. 
If fig. \ref{plot3} we show the ansatz for the potential for the state $(l,m)=(0,0)$ minus the isotropic potential for these parameters. 
\par

We study the dependence of the binding energies on the anisotropy parameters. 
We solve the 1-dimensional Schr\"odinger equation (\ref{my-sch}, \ref{my-sch2}) using the ansatz (\ref{ansatz1}), shifted so the threshold is set to zero, and with the screening mass $\tilde m_D$ replaced by the angle averaged masses (\ref{M-ave}). 
In our calculations we use $T=196$ MeV and $M_Q = 2m_b$ with $m_b = 4.676$ GeV. 
The Debye mass is $m_D=gT \sqrt{(N_c+N_f/2)/3}$ and we take $N_c=3$ and $N_f=3$.  

First we consider the ground state $(n,l,m)=(1,0,0)$. 
We use $g=1.85$ which is a typical value from the literature  \cite{Dumitru:2009ni,Islam:2022qmj}.  
We note this value is larger than one, which appears to contradict the assumption that a perturbative approach is valid. Recall that the coordinate space potential is obtained by Fourier transforming the  propagator in momentum space, and even though the polarization tensor is calculated perturbatively, the potential is explicitly non-perturbative in the coupling constant. This can be seen already from the isotropic calculation where the Fourier transform of the HTL propagator $-1/(p^2+m_D^2)$ gives the isotropic potential $-g^2 C_F/(4\pi r)e^{-r m_D}$ with $m_D \sim gT$. If we had expanded the propagator in $g$ we would have obtained the Coulomb potential $-g^2 C_F/(4\pi r)$,  which is just the leading term in the expansion in $r$ of the isotropic potential. The unexpanded isotropic potential is not a polynomial in $g$ and it correctly includes the thermal screening. These observations are typically used to justify the use of larger values of $g$.  

Table \ref{table3} shows the ground state binding energy minus the binding energy of the isotropic state for each set of parameters. The magnitude of the binding energy always increases, which means that anisotropy promotes binding in the ground state, and is consistent with the fact that the ansatz gives a deeper potential (see fig. \ref{plot3}).
\begin{table}[H]
\centering 
\begin{tabular}{|c|c|c|c|c|c|c|c|} 
\hline
 ~~~ ~~- ~~~~~&~~~ $ \xi_2$ ~~~&~~~ $ \xi_4$  ~~~&~~~ $ \xi_{14}$  ~~~&~~~ $ \xi_{11}$~~~ \\
\hline
 -0.617 & -4.30 & -1.52 & -1.52 & -0.866 \\
\hline 
\end{tabular}
\caption{Ground state binding energy minus the binding energy of the isotropic state. In all cases $\xi_9=0.95$ and $g=1.85$.  The parameter indicated in the top row is $0.8$ and all other parameters are set to zero. }
\label{table3}
\end{table}

A Yukawa potential of the form $-g^2 C_F e^{-m_D r}/(4\pi r)$ does not bind if the screening is too large. The criticallity condition depends on the value of the quantum number $l$ and has the form $2 a m_D <n_l$ where $a= 4\pi/(C_F g^2 M_Q)$ is the Bohr radius and $n_0 \approx 1.2$ and $n_1\approx 0.23$. These values agree with those of ref. \cite{yuk-chin} with the appropriate conversions of units and the mass parameter. Using our value of the temperature and quark mass we have $g_c\approx 0.41$ for states with $l=0$ and $g_c\approx 2.1$ for $l=1$. 
This means that we do not have bound states with $l=1$ using $g=1.85$. 
However, we have not included a string tension term in the ansatz that we use for the potential. 
This term would modify the long distance form of the potential so that bound states would form more easily. The purpose of our calculation is only to study the effect of anisotropy and we can do that without introducing a constant that must be fitted using information outside the scope of our  calculation by simply increasing the value of $g$. We have therefore also considered $g=2.85$, which is above the critical value for $l=1$. 

Table \ref{table4} gives the binding energy divided by the binding energy of the isotropic  state for each set of anisotropy parameters, for several different states,  with $g=2.85$. The results show that $\xi_{2}$ is particularly important, especially when $l=1$. 
In most cases the magnitude of the binding energy increases but for the state $(2,1,1)$ it can slightly decrease. 
\begin{table}[H]
\centering 
\begin{tabular}{|c|c|c|c|c|c|c|c|} 
\hline
~~state ~~& ~~~~~-~~~~~ & ~~~~$ \xi_2$ ~~~~&~~~~ $ \xi_4$ ~~~ ~&~~~~ $ \xi_{14}$ ~~~ ~&~~~~ $ \xi_{11}$~~~~ \\
\hline
(1,0,0) & 1.0041 &  1.0283 & 1.0100 & 1.0100 & 1.0057 \\
(2,0,0) & 1.0409 & 1.2866 & 1.1004 & 1.1004 & 1.0573 \\
(2,1,0) & 1.2477 & 1.5152 & 1.2844 & 1.4795 & 1.287 \\
(2,1,1) & 0.97462 & 1.4527 & 1.1052 & 1.0011 & 0.99507 \\
\hline 
\end{tabular}
\caption{Binding energy divided by the binding energy of the isotropic  state with $g=2.85$ and $\xi_9=0.95$. }
\label{table4}
\end{table}

\section{Conclusions}
\label{conclusions-sec}

We have calculated the real part of the heavy quark potential using hard-loop (HL) resummed perturbation theory with an anisotropic distribution function. This function depends on a set of anisotropy parameters that  can be chosen so that the momentum distribution provides a more realistic description of a quark-gluon plasma than has been used in previous calculations. 
We have derived an analytic expression for the heavy quark potential in a chirally asymmetric plasma, and found numerical solutions in the chirally symmetric limit. 
In the simple case of a spheroidally asymmetric plasma the potential becomes deeper (more strongly binding) as the anisotropy increases and the  quark-antiquark pair attracts more strongly when they are aligned with the direction of the anisotropy.
More realistic momentum distributions produce a much richer structure.

We have obtained an analytic expression for the potential in the limit of weak anisotropies. 
Using an ansatz and an averaging procedure we have constructed a potential that depends on one radial coordinate but still incorporates some of the anisotropy of the original system.  
We have used this 1-dimensional potential to study the dependence of the binding energy on the anisotropy parameters and found that the magnitude of the binding energy typically increases with anisotropy. This result is interesting because it means that the number of bound states that are formed at a given temperature will depend on the anisotropy of the momentum distribution. 

The imaginary part of potential can be calculated using the techniques we have developed in this paper. These results are particularly interesting because they will allow us to study the effect of anisotropy on the dissociation temperatures of quarkonium bound states. 

\newpage

\appendix

\section{The dressing functions for weak anisotropy to leading order in $p_0\to 0$}
\label{dress-ana}

In this section we give analytic results for the components of the dressing functions that we use to calculate the static potential. We consider only a chirally symmetric system and we work in the limit of weak anisotropy. In the limit $p_0\to 0$ the dressing functions can be expanded as in equation (\ref{df-exp}). For $C_\xi=1$ the coefficients of these expansions that are needed to calculate the heavy-quark potential are 
\begin{align}
\bar{\pi}_{1R}^{(0)}&=-\frac{1}{3} \xi _6 \sqrt{1-x^2} x \cos (\phi )-\frac{1}{15} \xi _8 \sqrt{1-x^2} x \cos (\phi ) \left(\left(2 x^2-5\right) \cos (2 \phi )+2 \left(x^2+2\right)\right)\nn\\
&+\frac{1}{15} \xi _{13} \left(4 x^2-3\right) \sqrt{1-x^2} x \cos (\phi )+\frac{1}{6} \xi _2 \left(\left(x^2-2\right) \cos (2 \phi )+x^2\right)+\frac{2}{15} \xi _{14} \left(2 x^2-3\right) x^2\nn\\
&-\frac{\xi _9 x^2}{3}+\frac{1}{30} \xi _4 \left(\left(4 x^4-2 x^2-8\right) \cos (2 \phi )+\left(x^4-5 x^2+4\right) \cos (4 \phi )+3 \left(x^4+x^2\right)\right)
\nn\\
&+\frac{1}{30} \xi _{11} \left(-4 x^4+x^2+\left(-4 x^4+7 x^2-2\right) \cos (2 \phi )\right)\,,\\
\bar{\pi}_{2R}^{(2)}&=-1+ \xi_0+\frac{1}{3} \xi _8 x \sqrt{1-x^2} \left(\cos (3 \phi )-4 x^2 \cos ^3(\phi )\right)+\frac{1}{6} \xi _2 \left(3 \left(x^2-1\right) \cos (2 \phi )+3 x^2+1\right)\nn\\
&-\xi _6 x \sqrt{1-x^2} \cos (\phi )+\frac{1}{3} \xi _{13} x \left(4 x^2-3\right) \sqrt{1-x^2} \cos (\phi )+\xi _9 \left(\frac{2}{3}-x^2\right)\nn\\
&+\frac{1}{30} \xi _4 \left(15 x^4+5 \left(x^2-1\right)^2 \cos (4 \phi )+10 \left(2 x^4-x^2-1\right) \cos (2 \phi )+3\right)\nn\\
&+\frac{1}{30} \xi _{11} \left(-20 x^4+15 x^2-5 \left(4 x^4-5 x^2+1\right) \cos (2 \phi )+1\right)\nn\\
&+\frac{1}{15} \xi _{14} \left(20 x^4-30 x^2+9\right)\,,\\
\bar{\pi}_{3R}^{(0)}&=\frac{1}{3} \xi _2 \left(2 x^2-1\right) \cos ^2(\phi )-\frac{2}{3} \xi _6 x \sqrt{1-x^2} \cos (\phi )+\frac{2}{15} \xi _{14} \left(8 x^4-12 x^2+3\right)\nn\\
 &-\frac{1}{15} \xi _8 x \sqrt{1-x^2} \cos (\phi ) \left(\left(8 x^2-5\right) \cos (2 \phi )+8 x^2+1\right)\nn\\
 &+\frac{4}{15} \xi _{13} x \left(4 x^2-3\right) \sqrt{1-x^2} \cos (\phi )+\xi _9 \left(\frac{1}{3}-\frac{2 x^2}{3}\right)\nn\\
 &+\frac{2}{15} \xi _4 \cos ^2(\phi ) \left(4 x^4+x^2+\left(4 x^4-5 x^2+1\right) \cos (2 \phi )-2\right)\nn\\
 &+\frac{1}{30} \xi _{11} \left(2 \left(1-2 x^2\right) \sin ^2(\phi )-4 \left(8 x^4-8 x^2+1\right) \cos ^2(\phi )\right)\,,\\
\bar{\pi}_{6R}^{(0)}&=\frac{1}{6} \xi _6 \sqrt{1-x^2} \sin (\phi )+\frac{1}{30} \xi _{11} x \left(6 x^2-5\right) \sin (2 \phi )+\frac{1}{10} \xi _{13} \left(1-2 x^2\right) \sqrt{1-x^2} \sin (\phi )\nn\\
&-\frac{1}{5} \xi _4 x \sin (2 \phi ) \left(\left(x^2-1\right) \cos (2 \phi )+x^2\right)+\frac{1}{10} \xi _8 \sqrt{1-x^2} \sin (\phi ) \left(\left(3 x^2-1\right) \cos (2 \phi )+3 x^2\right)\nn\\
&-\frac{1}{3} \xi _2 x \sin (\phi ) \cos (\phi )\,.
\label{dress-ana-real}
\end{align}
\begin{align}
  \bar{\pi}_{4I}^{(1)}&=\frac{1}{4} \pi  \xi _2 x \sqrt{1-x^2} \cos ^2(\phi )+\frac{3}{16} \pi  \xi _4 x \sqrt{1-x^2} \cos ^2(\phi ) \left(\left(x^2-1\right) \cos (2 \phi )+x^2+1\right)\nn\\
&+\frac{1}{8} \pi  \xi _6 \left(2 x^2-1\right) \cos (\phi )+\frac{1}{16} \pi  \xi _{11} x \sqrt{1-x^2} \left(\left(2-3 x^2\right) \cos (2 \phi )-3 x^2+1\right)\nn\\
&+\frac{3}{32} \pi  \xi _8 \left(x^2 \left(4 x^2-3\right) \cos ^3(\phi )+\left(2 x^2-1\right) \sin ^2(\phi ) \cos (\phi )\right)-\frac{1}{8} 3 \pi  \xi _{14} x \left(1-x^2\right)^{3/2}\nn\\
&-\frac{1}{4} \pi  \xi _9 x \sqrt{1-x^2}-\frac{3}{32} \pi  \xi _{13} \left(4 x^4-5 x^2+1\right) \cos (\phi )\,,\\
\bar{\pi}_{5I}^{(1)}&=\frac{1}{32} \pi  \xi _{11} \left(3 x^2-1\right) \sqrt{1-x^2} \sin (2 \phi )+\frac{3}{32} \pi  \xi _{13} x \left(x^2-1\right) \sin (\phi )\nn\\
&-\frac{3}{32} \pi  \xi _8 x \sin (\phi ) \left(\left(3 x^2-2\right) \cos ^2(\phi )+\sin ^2(\phi )\right)-\frac{1}{4} \pi  \xi _2 \sqrt{1-x^2} \sin (\phi ) \cos (\phi )\nn\\
&-\frac{3}{32} \pi  \xi _4 \sqrt{1-x^2} \sin (2 \phi ) \left(\left(x^2-1\right) \cos (2 \phi )+x^2+1\right)-\frac{1}{8} \pi  \xi _6 x \sin (\phi ) \,. 
\label{dress-ana-imag}
\end{align}

\end{document}